\documentclass[aps,prl,amsmath,amssymb,reprint]{revtex4-2}
\usepackage{lmodern}
\usepackage{graphicx}
\usepackage{dcolumn}
\usepackage{bm}
\usepackage[utf8]{inputenc}
\usepackage[T1]{fontenc}
\usepackage{mathptmx}
\usepackage{hyperref}
\usepackage{etoolbox}

\usepackage[english]{babel}  
\DeclareMathOperator*{\argmin}{argmin}

\begin{document}
\title{Data-driven complete basis set limit estimates from a minimal auxiliary basis}
\author{Nicolas Grimblat}
\affiliation{Institut f\"ur Chemie, Universit\"at Kassel, Heinrich-Plett-Stra{\ss}e~40, 34132~Kassel,~Germany}
\affiliation{Center for Interdisciplinary Nanostructure Science and Technology (CINSaT), Heinrich-Plett-Stra{\ss}e 40, 34132 Kassel}
\author{Gabriel Klassen}
\affiliation{Institut f\"ur Chemie, Universit\"at Kassel, Heinrich-Plett-Stra{\ss}e~40, 34132~Kassel,~Germany}
\affiliation{Center for Interdisciplinary Nanostructure Science and Technology (CINSaT), Heinrich-Plett-Stra{\ss}e 40, 34132 Kassel}
\author{Guido Falk von Rudorff}%
\email{vonrudorff@uni-kassel.de}
\affiliation{Institut f\"ur Chemie, Universit\"at Kassel, Heinrich-Plett-Stra{\ss}e~40, 34132~Kassel,~Germany}
\affiliation{Center for Interdisciplinary Nanostructure Science and Technology (CINSaT), Heinrich-Plett-Stra{\ss}e 40, 34132 Kassel}

\date{\today}
\begin{abstract}
Quantum chemistry calculations are often performed using atom-centered basis sets which are chosen to balance accuracy and cost. While they are systematically improvable, the total energy converges slowly with basis set size towards the complete basis set (CBS) limit. Common extrapolation methods require several intermediate-quality calculations to afford an estimate of the CBS energy. We propose combining a pairwise interaction model with a minimal complementary auxiliary basis set (CABS) baseline to estimate the CBS energy from a single quantum chemistry calculation in a minimal basis set via Kernel-Ridge-Regression (KRR), which is more efficient than both direct and $\Delta$-machine learning. We show that KRR on standard molecular representations can be improved by approximating atom-wise local kernels using Chebyshev polynomials which allows us to train KRR models efficiently on moderate compute resources, further enabling a data-driven approach towards CBS combining physical baselines capturing leading order effects with data-efficient machine learning models.
\end{abstract}

\maketitle

\section{Introduction}
The accurate prediction of molecular energies across broad chemical space still depends critically on controlling basis-set incompleteness error (BSIE). In practice, many workflows require balancing computational affordability against proximity to the complete basis set (CBS) limit. 
Correlation-consistent basis sets (e.g. cc-pVXZ, X=$D, T, Q, 5, 6$) were designed to recover correlation energy in a near-systematic fashion, enabling extrapolation to the CBS limit\cite{Dunning1989}, e.g. with inverse‑power laws of the form
\begin{equation}
 E_\mathrm{corr}^X  = E_\mathrm{corr}^{CBS} + A X^{-3}
 \label{eqn:helgaker}
\end{equation}
    where $X$ is the cardinal number \cite{Helgaker1997}. This concept has been widely adopted and generalized for other systems and methods \cite{Xi2024}. Systematic analyses of Hartree–Fock and correlated convergence have shown regular asymptotic trends as well as method-dependent deviations\cite{Halkier1998,Helgaker1997}. For this reason, extrapolation formula choice and cardinal-pair selection remain critical in practical workflows\cite{Schwenke2005,Parameswaran2024,Lang2025}, but their reliability can degrade when the selected basis-set pair is not fully in the asymptotic convergence regime.  In general, the total energy at the cardinal number $X$, can be written as
\begin{equation}
 E(X)  = E_\mathrm{CBS} + A f(X)
 \label{eqn:general2point}
\end{equation}
where $E_\mathrm{CBS}$ is the CBS limit, $A$ is a system-dependent coefficient, and $f(X)$ describes the decay of the BSIE. Different such decay functions have been proposed and evaluated both for correlated electrons\cite{Martin1996,Truhlar1998,Huh2003,Bakowies2007,Bakowies2007a,Varandas2000} and for mean-field method extrapolation\cite{Truhlar1998,Klopper1986,Feller1992,Jensen2001,Karton2006,Petersson1988,Martin1997,Halkier1998,Jensen1999,Jensen2005,Kraus2020}, where practical results often prefer a different decay function than theoretical motivation suggests\cite{Schwenke2005}. Using a linear combination of two calculations with different basis sets at cardinal numbers $X$ and $Y$, elimination of $A$ yields a two-point extrapolation framework, e.g. for correlation energies 
\begin{align}
    E_\mathrm{corr}^\mathrm{CBS} =     \frac{X^{\beta} E_{\mathrm{corr}}(X) - Y^{\beta} E_{\mathrm{corr}}(Y)}{X^{\beta} - Y^{\beta}}
    \label{eq:truhlar-schwenke}
\end{align}
where $f(X)=X^{-\beta}$ controls the rate of the asymptotic convergence. While $\beta=3$ is often assumed in the asymptotic regime, several studies found deviations for lower cardinal numbers\cite{Truhlar1998,Neese2011}. Recent benchmarks further emphasize that these exponents are not universal constants, but depend on both method and basis set family, such that re-optimization for each pair can significantly improve accuracy across the chemical space\cite{Parameswaran2024,Lang2025}.

For Hartree-Fock (HF) energies, convergence with respect to basis set size is typically faster and better described by an exponential-square-root form,
\begin{equation}
    E_\mathrm{HF}(X) = E_\mathrm{HF}^\mathrm{CBS} + A\exp\left(-{\alpha}{\sqrt{X}}\right)
    \label{eq:neese}
\end{equation}
where $\alpha$ governs residual HF BSIE decay. In practice $\alpha$ and $\beta$ are effective parameters that may be optimized empirically for a given basis-set.

By contrast, F12 methods are explicitly correlated and instead address the slow convergence at its origin, the electron–electron cusp. By incorporating interelectronic-distance dependence directly, F12 methods dramatically accelerate convergence relative to conventional expansions and can recover near-CBS energies with substantially smaller basis sets\cite{Lang2025,Helgaker1997,Peterson2008,Adler2007,Knizia2009}, often with two cardinal numbers less\cite{Kumar2020,Adler2007}, owing to improved $X^{-7}$ scaling\cite{Kutzelnigg1991}.
F12 requires electron correlation to be included in the underlying quantum chemistry method, so it is not directly applicable to Hartree-Fock, but can be extended to address it using the complementary auxiliary basis set (CABS) singles term\cite{Valeev2004} which will be used to build a physics-informed but data-driven correction in this work. CABS/OptRI residuals have large signed errors\cite{Shaw2017} which benefit from an atomic correction. Other formalisms pursue variants which can be applied to HF and DFT alike\cite{Liang2004,Wolinski2003}.

Modern machine learning (ML) approaches for this extrapolation problem often adopt an atomic decomposition of total energies, where each atomic contribution depends on the local environment both for direct learning of the total energy \cite{Unke2019,Schutt2018,Holm2023,Speckhard2025,Qiao2020,Christensen2021,Qiao2022,Batzner2022} or for only considering the CBS correction\cite{Otero-de-la-Roza2017,Otero-de-la-Roza2020,Prasad2022,Prasad2024,Qu2021}.  For the CBS limit of correlated methods, this typically involves learning extrapolated energies\cite{Smith2019}. Alternative approaches focus on proposing new interactions\cite{Kruse2012} or analysis of the emergent interactions\cite{Esders2025} yielding a correlated complete basis set result.

The hierarchical structure of the problem naturally supports detrending: a physics‑based baseline energy model (e.g., low‑cost semiempirical, DFT, or approximate CCSD(T)/finite‑basis result) is decomposed into atomic or local contributions, and the ML model learns a residual correction, commonly referred to as $\Delta$-ML\cite{Ramakrishnan2015}. This exploits that the learning curve (i.e. the expected error as function of training data points $N$) scales with the standard deviation of the labels $\sigma_y$: reducing the standard deviation by only considering the difference to a baseline shifts the learning curve, effectively reducing the total data points needed for learning the same quantity at the added cost of the baseline evaluation. The generalization thereof to problems with exact hierarchy of basis sets and levels of theory is multi-fidelity learning\cite{Vinod2023,Vinod2025,Zaspel2018} which builds linear combinations of individual sub-models akin to an ensemble estimator.

Reducing the basis set incompleteness error is relevant\cite{Morgan2018,Karton2019,Drabik2024} for many methods relying on or calculating forces and other gradients such as geometry optimization and molecular dynamics\cite{Pathak2023,Ruiz-Serrano2012}, differentiable chemistry\cite{Kasim2022,TamayoMendoza2018,Abbott2021,Zhang2022}, quantum alchemy\cite{Domenichini2020,Domenichini2024,Rudorff2021a,Lilienfeld2009}, response functions and perturbation theory\cite{Brakestad2020,Hurtado2024,Traore2022}.

In the long history of approximations in quantum chemistry, elements naturally have been treated as distinct entities, for example, in the design of pseudo potentials or basis functions. This is a particular issue for mixed response functions, that is where geometric changes are coupled with electronic changes\cite{Rudorff2024}. Since the Hellmann-Feynman theorem, which conceptually offers a particularly convenient calculation of electronic gradients, only holds if the basis set does not change in the direction of that derivative, Pulay correction terms have to be included for incomplete basis sets. In this work, we focus on Hartree-Fock CBS, since it is the dominating physical component to corrections of higher level methods and published databases offer a wealth of data which allows for a more principled analysis for this method.

\begin{figure}[h]
    \centering
    \includegraphics[width=1\linewidth]{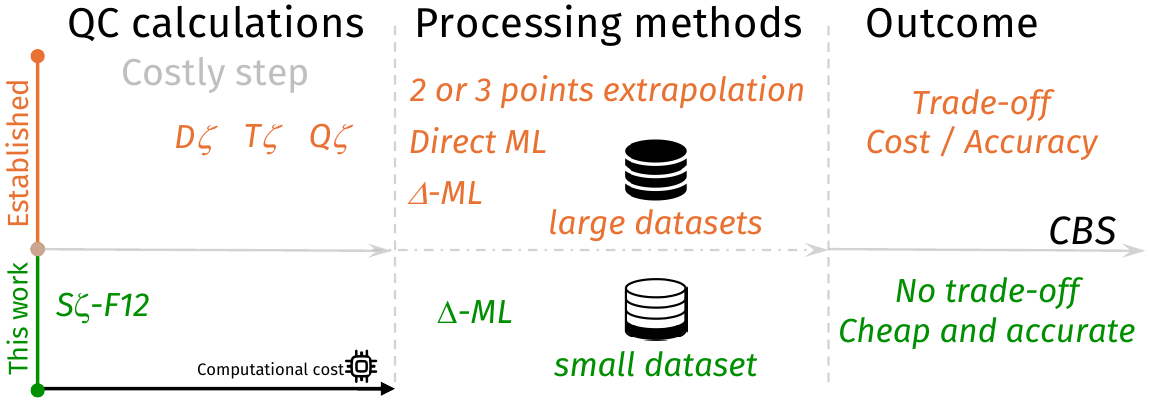}
    \caption{Comparison of workflows towards the estimation of the complete basis set limit. On top, in orange, traditional and established approaches to predict the energy value. On the bottom, in green, the methodology developed in this work.}
    \label{fig:placeholder}
\end{figure}

\section{Methods}
Instead of learning the difference between some medium and high quality calculation, we instead deliberately reduce the basis set quality of the baseline quantum chemistry calculation and at the same time add  a physical correction, the Complementary Auxiliary Basis Set (CABS) correction, in order to explain part of that change. We
  exploit computational efficiency by stripping down and optimizing a CABS basis set for this particular application. This allows us to make the
  baseline calculation substantially cheaper while
  at the same time improving prediction outcome. This offers a systematically improvable way to reduce prediction time, and thus, the total time to result.

\subsection{Kernel Ridge Regression}
For a molecule of given nuclear charges $Z_I$ and positions $\mathbf{R}_I$, one can find a representation $x(Z_I, \mathbf{R}_I) \in \mathbb{R}^d$ for each of $n$ molecules ($X\in \mathbb{R}^{n\times d}$). With those representations, we employ Kernel-Ridge-Regression (KRR) where a kernel function $k(x, x')$ acts as similarity measure in chemical space. 
\begin{align}
    f(x_q) = \sum_{i=1}^{n_t} \alpha_i k^G(x_q, x_i)\label{eqn:globalinterpolant}
\end{align}
where $x_q$ is the representation of the query molecule and the model coefficients $\alpha$ are found by minimizing the loss function
\begin{align}
    \mathcal{L} = \sum_i^{n_t} (f(x_i)-y_i)^2 + \lambda \|f\|^2\qquad;\qquad \|f\|^2\equiv\boldsymbol{\alpha}^T\mathbf{K}\boldsymbol{\alpha}\label{eqn:loss}
\end{align}
with the Kernel matrix element $\mathbf{K}_{ij} = k^G(x_i, x_j)$ and regularization strength $\lambda$. The model for a \textit{fixed} representation $x(Z_I, \mathbf{R}_I)$ can be obtained directly via
\begin{align}
    \boldsymbol{\alpha} = (\mathbf{K}+\lambda \mathbf{I})^{-1}\mathbf{y}
    \label{eqn:krr}
\end{align}

The coefficients $\boldsymbol{\alpha}$ implicitly depend on the regularization parameter $\lambda$ and on the internal parameters of the kernel function $k$, which together are called hyperparameters. To find the optimal combination of hyperparameters $\boldsymbol{\theta}$ which minimize the expected value of the loss function, the model \ref{eqn:krr} is built with different combinations of hyperparameters during cross validation. In this work, we employ nested Monte-Carlo cross validation where a full dataset $S$ is repeatedly partitioned into $p$ random disjoint subsets $T_i$ (training) with training size $N=|T_i|$ and $H_i$ (holdout). For each such split, the training part $T_i$ is further partitioned randomly into $s$ shuffles of training data $T_{ij}^T$ and the corresponding validation set $T_{ij}^V$, such that $T_{ij}^V\cup T_{ij}^T = T_i$ for all shuffles $j$. Since the loss function eqn.~\ref{eqn:loss} contains the mean squared error (MSE), we choose the root mean squared error (RMSE) as unit-compatible target metric. We will denote the RMSE for a model trained on dataset $A$ and predicted on $B$ as $\mathcal{E}(A\rightarrow B|k, \boldsymbol{\theta})$.

For each partition $i$, this yields a validation error
\begin{align}
    \mathcal{E}(T_i|k, \boldsymbol{\theta})=\frac{1}{s}\sum_j \mathcal{E}(T_{ij}^T\rightarrow T_{ij}^V|k, \boldsymbol{\theta})
\end{align}
which is minimized to obtain the optimal hyperparameters
\begin{align}
    \boldsymbol{\theta}_i = \argmin_{\boldsymbol{\theta}}\ \mathcal{E}(T_i|k, \boldsymbol{\theta})
\end{align}
The expected model accuracy on unseen data (the generalization error) $\mathcal{E}_k(S)$ then is 
\begin{align}
    \mathcal{E}_k(S, N) = \frac{1}{p}\sum_i \mathcal{E}(T_i \rightarrow H_i|k,\boldsymbol{\theta}_i)
\end{align}
This particular method avoids bias from a single particular split, and yields a reliable estimate of the underlying generalization error in particular in the low-data regime which is of high practical relevance. The random permutations allow for dynamic dataset-specific sampling of splits until the results are converged w.r.t. number of permutations $p$ and splits $s$. While slightly more expensive than $k$-fold cross validation, it is systematically improvable by keeping the  validation set size $|T_{ij}^V|$ independent from the number of splits $s$. We find that the test error closely matches the validation error as expected: $\mathcal{E}(T_i|k, \boldsymbol{\theta}_i)\approx \mathcal{E}(T_i \rightarrow H_i|k,\boldsymbol{\theta}_i)$.

We speed up and stabilize our hyperparameter optimization using the Schur complement, the eigendecomposition, by using Kernel centering and Chebyshev polynomials (see below).

\subsection{Schur Complement}
Naive implementation of eqn.~\ref{eqn:krr} for each model in the cross validation requires either a Cholesky decomposition or a matrix inversion, both $\mathcal{O}(N^3)$. For training, this is the dominant cost, since typically thousands of models are trained to find the correct hyperparameters. While the Cholesky decomposition is numerically more stable, the inversion allows for a particularly efficient implementation where the marginal cost of additional shuffles is much lower (often by two orders of magnitude) by using the Schur complement.

First we calculate the matrix $\mathbf{M}$ which is the Kernel matrix of the training and validation data, the size of $\mathbf{M}$ is $(N_q+N_p)\times (N_q+N_p)$, where $N_p$ is the number of validation points and $N_q$ is the number of training points. This yields the $2\times 2$ blockmatrix
\begin{align}
    \mathbf{M} = \begin{bmatrix}
\mathbf{K}_{tt} & \mathbf{K}_{tv} \\
\mathbf{K}_{vt} & \mathbf{K}_{vv}
\end{bmatrix}
\end{align}
The submatrices have the size $\mathbf{K}_{tt} (N_q\times N_q)$, $\mathbf{K}_{tv} (N_q\times N_p)$, $\mathbf{K}_{vt} (N_p\times N_q)$ and $\mathbf{K}_{vv} (N_p\times N_p)$. The inversion of $\mathbf{M}$ yields another $2\times 2$ blockmatrix 
\begin{align}
    \mathbf{M}^{-1} = \begin{bmatrix}
\mathbf{E} & \mathbf{F} \\
\mathbf{G} & \mathbf{H}
\end{bmatrix}
\end{align}
the sizes of the submatrices will be conserved during the inversion. Since $\mathbf{M}$ is larger than $\textbf{K}_{tt}$ this initial inversion comes with a higher cost than the inversion of the matrix $\textbf{K}_{tt}$. But for further shuffle splits, the matrices $\mathbf{M}$ and $\mathbf{M}^{-1}$ can be shuffled jointly using the permutation matrix $\mathbf{P}$
\begin{equation}
    \mathbf{M}' \equiv \mathbf{P}\mathbf{M}\mathbf{P}^{\text{T}}\Rightarrow  \mathbf{M}'^{-1} \equiv \mathbf{P}\mathbf{M}^{-1}\mathbf{P}^{\text{T}}
\end{equation}
Therefore, it suffices to calculate $\mathbf{M}$ and $\mathbf{M}^{-1}$ once and reuse it for different splits during the cross-validation.

The inversion of a $2\times 2$ blockmatrix can be calculated by using the Schur complement, applied on $\mathbf{M}^{-1}$, which yields\cite{Lu2002}:
\begin{equation}
 \mathbf{K}_{tt}= (\mathbf{E}-\mathbf{F}\mathbf{H}^{-1}\textbf{G})^{-1}\Rightarrow \mathbf{K}_{tt}^{-1}=\mathbf{E}-\mathbf{F}\mathbf{H}^{-1}\mathbf{G}
 \label{eqn:Schur}
\end{equation}
Therefore, the inversion of $\mathbf{K}_{tt}$ can be substituted with the inversion of $\mathbf{H}$, which has the size $(N_p\times N_p)$. With an asymptotical scaling of $\mathcal{O}(N^3)$ for matrix inversion and a standard test/train-split of $20/80$, the computational cost is reduced to $1/4^{3}\approx 1.6\%$ although scaling remains the same.
\subsection{Eigendecomposition}
The kernel matrix $\mathbf{K}$ depends only on the parameters of the kernel function for a fixed data set. For this reason, it might be beneficial to calculate the kernel matrix for one set of initial parameters and use the eigendecomposition of $\mathbf{K}$ to evaluate the performance for different values of the regularization parameter $\lambda$. The eigendecomposition of $\mathbf{K}$ is:
\begin{equation}
    \mathbf{K}= \mathbf{Q}\boldsymbol{\Lambda} \mathbf{Q}^{-1}
    \label{eqn:decomp}
\end{equation}
Here $\mathbf{Q}$ is the matrix consisting of the eigenvectors of $\mathbf{K}$ and $\boldsymbol{\Lambda}$ is a diagonal matrix with the eigenvalues $\lambda_i$ as its elements. Substituting $\mathbf{K}$ in eqn.~\ref{eqn:krr} with \ref{eqn:decomp} results in:
\begin{equation}
    \mathbf{Q}(\boldsymbol{\Lambda}+\lambda \mathbf{I})^{-1}\mathbf{Q}^\text{T} = \mathbf{Q}\mathbf{L}\mathbf{Q}^\text{T}\quad;\quad \mathbf{L} = \text{diag}\left(\frac{1}{\lambda_i+\lambda}\right)
\end{equation}
Hereby we substitute a matrix inversion with two matrix multiplications. While the multiplication with a diagonal matrix scales with $\mathcal{O}(N^2)$, the second multiplication scales with $\mathcal{O}(N^3)$ and, therefore, has the same formal scaling as a matrix inversion. 
We apply the eigendecomposition only together with the Schur complement. By inserting \ref{eqn:Schur} into \ref{eqn:krr} we obtain
\begin{equation}
    \alpha = (\mathbf{E}-\mathbf{FH}^{-1}\mathbf{G})\mathbf{y}=\mathbf{Ey}-(\mathbf{F}(\mathbf{H}^{-1}(\mathbf{Gy})))
    \label{eqn:eigenschur}
\end{equation}
The submatrices can be expressed by the Eigendecomposition of the Matrix $\mathbf{M}$ 
\begin{align}
\mathbf{E} &= \mathbf{Q}_\text{Train}\mathbf{L}\mathbf{Q}_\text{Train}^\text{T} &;\quad 
&\mathbf{F} = \mathbf{Q}_\text{Train}\mathbf{L}\mathbf{Q}_\text{Val}^\text{T}\nonumber\\
\mathbf{H} &= \mathbf{Q}_\text{Val}\mathbf{L}\mathbf{Q}_\text{Val}^\text{T}   &;\quad
&\mathbf{G} = \mathbf{Q}_\text{Val}\mathbf{L}\mathbf{Q}_\text{Train}^\text{T}
\end{align}
$\mathbf{Q}_\text{Train}$ and $\mathbf{Q}_\text{Val}$ are the truncated versions of $\mathbf{Q}$ from \ref{eqn:decomp} that only contain the rows of training points or, respectively, validation points. Using these expressions in \ref{eqn:eigenschur} allows us to write the equation \ref{eqn:krr} as a chain of matrix vector products, which scale with $\mathcal{O}(N^2)$, rather than calculating the matrix inversion. The eigendecomposition of $\mathbf{M}$ also allows us to skip the inversion of $\mathbf{M}$. 


\subsection{Kernel centering}
The standard approach in KRR is to center the labels, by subtracting the mean of the training data. This ensures that the prediction model is not biased by any underlying trend and (in Gaussian Process Regression terminology) aligns the expected prediction with the zero mean function that KRR implies. However, the features of the training data are generally not centered around the origin of the feature space. By using a kernel method like KRR to predict the label of a query molecule as shown in \ref{eqn:globalinterpolant} the model must map the uncentered representations in the feature space onto the centered labels. To achieve this, the model needs to account for the offset. Therefore, we center our kernel matrix following the expression\cite{Scholkopf1998}:
\begin{align}
    \mathbf{K'} 
    &=(\mathbf{K}-1_N\mathbf{K}-\mathbf{K}1_N+1_N\mathbf{K}1_N)
\end{align}
where $1_N$ is the matrix with all entries equal to $1/N$. By centering the kernel matrix, the largest eigenvalue is driven to zero, or numerically close to it, rendering the matrix rank deficient. Although a rank deficient matrix cannot be inverted directly, this is not an issue for KRR, since the inversion is not applied to K alone but to the regularized sum given in \ref{eqn:krr}. The eigenvalue zeroed out by centering is therefore shifted by $\lambda$, restoring the matrix to full rank. For the data and kernel functions used in this work, the dominant eigenvalue in the uncentered kernel originated from the constant offset. Centering removes this offset, thereby reducing the condition number. A better-conditioned matrix in turn enables the use of decomposition methods such as the Schur or eigendecomposition for a larger range of model parameters.

\subsection{Chebyshev polynomials for local kernels}
For some properties, a local kernel is advantageous where each atom is assigned its own representation $x_i^l$ or, equivalently, the full molecule is assigned a matrix $x_i$. This forms non-normalized kernel functions $k^U$ from atomic kernel functions $k^A$:
\begin{align}
    k^U(x_i, x_j) = \sum_{l=1}^{N_A}\sum_{m=1}^{N_B}\delta_{Z_l,Z_m}k^A(x_i^l, x_j^m)\label{eqn:localkernel}
\end{align}
with the optional Kronecker-Delta $\delta$ which is 1 if the nuclear charges $Z_l$ and $Z_m$ of the $N_A$ atoms are identical and zero otherwise which restricts comparison of atomic environments within each element only, suppressing alchemical interactions even if such elemental similarity might exist\cite{Faber2018}. If the Kronecker-Delta is included, we will refer to this as \textit{elemental} kernel, otherwise we will call it a \textit{local} kernel only.

Through normalization, one obtains the local equivalent $k^L$ to the global kernel $k^G$ in eqn.~\ref{eqn:globalinterpolant}:
\begin{align}
    k^L(x_i, x_j) = \frac{k^U(x_i, x_j)}{\sqrt{k^U(x_i, x_i)k^U(x_j, x_j)}}
\end{align}

During cross-validation, the hyperparameters of the kernel function will be changed iteratively and thereby trigger a full reevaluation of the kernel function on the pairwise distances between all training atoms. The cost of calculating a single kernel matrix element scales quadratically with the average number of atoms in the molecules for a local representation while it is of constant cost for global representations. Since for a radial basis function kernel such as the exponential or the Laplacian kernel, the sum of exponential functions cannot be simplified by just algebraic operations, it would be desirable to have a way of saving at least the unfavorably scaling way of double sum over the pairwise atoms. We obtain constant cost for a single kernel matrix element, even for a local representation with an exponential or Laplacian kernel, by approximating the exponential function over the domain $[-20,0]$ as Chebyshev polynomial over the domain $[-1,1]$. For $x>20$, $\exp(-x)$ is essentially 0. We reach a maximum absolute error to the exponential over the domain of about $10^{-9}$ with a polynomial of degree $d=20$. Approximating $\exp(x) \approx \sum_{p=0}^d c_px^p$, we can write e.g. the local exponential kernel in eqn.~\ref{eqn:localkernel} as 
\begin{align}
    k^U(x_i, x_j) &= \sum_{l=1}^{N_A}\sum_{m=1}^{N_B}\delta_{Z_l,Z_m}\exp(-\|x_i^l- x_j^m\|/\sigma)\\
    &\approx \sum_{l=1}^{N_A}\sum_{m=1}^{N_B}\delta_{Z_l,Z_m}\sum_{p=0}^dc_p\left[\frac{-\|x_i^l- x_j^m\|}{\sigma}\right]^p\\
    &= \sum_{p=0}^d \frac{c_p}{\sigma^p} \sum_{l=1}^{N_A}\sum_{m=1}^{N_B}\delta_{Z_l,Z_m} \left(-\|x_i^l- x_j^m\|\right)^p\\
    &=\sum_{p=0}^d \frac{c_p}{\sigma^p} C_{ijp}
\end{align}
with the kernel function length scale $\sigma$. This way, the double sum over the pairwise combinations of atoms can be replaced with a precomputed third-order tensor $C_{ijp}$ and the evaluation of a polynomial which can be done in constant time and memory regardless of the number of atoms. The same approximation can be made for the Gaussian kernel by inserting the squared distance and length scale.

\subsection{Complementary Auxiliary Basis Set Optimisation}
Expecting that the auxiliary basis set can be chosen to explain the non-trivial parts, we optimize a stripped-down CABS basis by choosing a particular fully decontracted set of
   shells with the exponents chosen to minimize the mean absolute error (MAE) after detrending. This allows
    the CABS correction to put all basis set flexibility into the non-trivial components. For Hydrogen atoms, we always only use S functions, while for heavy atoms, we explored other combinations. We calculate the gradients of the MAE with forward finite differences and use steepest descent
  with a line scan on a logarithmic grid. We detrend in the high-bias setting by applying a linear regression only on either stoichiometry alone (atomic detrending) or with a molecular pair interaction feature inspired by gCP\cite{Kruse2012} (pair detrending):
  \begin{align}
      E_\textrm{pair} = \sum_{i<j} Z_i^p Z_j^p  \exp(-\theta R_{ij})
  \end{align}
  with nuclear charges $Z_i$ and interatomic distance $R_{ij}$. We choose $p = 2.5$ and $\theta=3$, since the fit shows a wide parameter range with equivalent results. For added flexibility, measure the exact timing and to support our optimization procedure, we reimplemented the CABS correction as implemented in the MP2-F12 routines of Psi4\cite{Mitchell2024} in Python using PySCF\cite{Sun2017} and AI assistance and verified the equivalence of the results.

\section{Results}
We selected the GDB-BSIE dataset\cite{Holm2023}, which was designed for CBS extrapolation studies, including sequential energy values from increasing cardinal points ranging from STO-3G to cc-pV5Z. The largest basis set in this hierarchy, 5Z, achieves an accuracy close to the CBS limit, within possible computational cost constraints considering its size, making it our target energy. While this leaves a residual to the complete basis set limit, the difference between QZ and 5Z after detrending is less than 1\,mHa and the difference between 5Z and 6Z is expected to be even smaller. The GDB-BSIE dataset contains ten batches of molecular configurations, spanning different conformers. Since we found batch nine thereof to mistakenly report B3LYP total energies in the HF dataset, we excluded it from all our considerations.

\begin{figure}
    \centering
    \includegraphics[width=\linewidth]{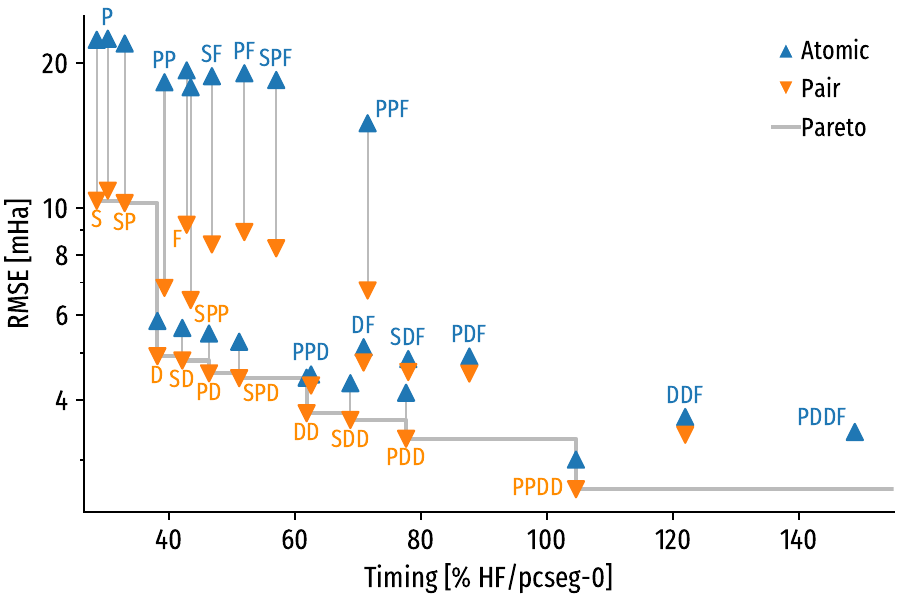}
    \caption{Residual nonlinear error as function of the computational cost of the optimized CABS correction relative to the HF/pcseg-0 single point for atomic detrending and for pair detrending. The pareto frontier connects the points forming the optimal tradeoff. The labels denote the different CABS shells for all heavy elements (Hydrogen always has one S shell), e.g. \textit{PDD} denotes a CABS basis of one P and two D shells.}
    \label{fig:cabsopt}
\end{figure}

\subsection{CABS optimization}
For our optimization of the CABS basis, we use 200 random molecules from the GDB-BSIE data set. They have been stratified to span the full elemental diversity of the dataset: C, H, N, O, F, S, Cl.
  Since this selection is particularly small and typically allows only a few parameters per element, we do not expect noticeable overfitting.  The timing information has been obtained by comparing the evaluation cost of the Hartree-Fock single point and the additional steps necessary to evaluate the CABS correction using our PySCF implementation. 
  
  Figure~\ref{fig:cabsopt} shows that the key component for an effective and cheap CABS correction over plain pcseg-0 calculations is adding d functions, as almost all points on the pareto front contain d functions. Moreover, the introduction of d functions in the first place reduces the RMSE after detrending and CABS correction to about 5\,mHa. That matches the common estimate that higher cardinal number basis sets each introduce basis functions of the next higher angular momentum. Since pcseg-0 contains s and p functions, the introduction of d functions is expected to be  helpful. The optimization itself is well-conditioned, since no CABS basis that is a strict superset of a smaller basis set yields a lower error. It is interesting, however, that the combination of d and f functions can not improve significantly over d functions alone.  
  We find that optimizing the CABS exponents without removing the linear trend of atomic contributions is not efficient, suggesting that the CABS correction on its own covers mostly a linear trend in atomic contributions in a comparably costly manner. With detrending however, CABS applied to the minimal basis pcseg-0 reaches cc-pVDZ accuracy at a fraction of the cost.

  We considered two variants of the detrending: one which only takes into account the stoichiometry and one which includes pair detrending inspired by gCP (see above). We observe that the effect of the pair detrending is strongest if no d functions are included in the CABS and that no pair detrending reaches the accuracy of any form of inclusion of d functions alone. This suggests that the gCP pairwise contributions are mostly of d character but do not explain $d$ effects on their own. Pair detrending is most effective for small CABS -- as it only contains a single fitting parameter, this approach can drastically reduce cost, e.g. pair-detrended CABS with a single d function is about as accurate as atom-detrended CABS with an additional f function at about half the computational cost.

  The CABS we considered were substantially smaller than common approaches such as AutoCABS which is essential to reduce the computational cost. While adding more functions adds degrees of freedom which can reduce the CABS error further, Fig.~\ref{fig:cabsopt} clearly shows that this improvement convergence is slow with the maximum angular moment. Since higher angular momenta become increasingly costly, since each adds $2l+1$ basis functions. We therefore suggest to use CABS for CBS estimation only in conjunction with atomic or pair detrending and with $d$ functions to save substantial
   computational cost. In this work, this approach provides a highly efficient baseline for machine learning methods (see below).

\subsection{Accelerated Kernel Ridge Regression}
Using the combination of kernel centering, Schur and eigendecomposition in general, and for local representations Chebyshev polynomials, we achieve a 10x-50x speedup compared to  canonical KRR, each on a single core. The vast speed up in training models makes it possible to run cross-validation over a comprehensive grid of hyperparameters. 

\begin{figure}
    \centering
    \includegraphics[width=\linewidth]{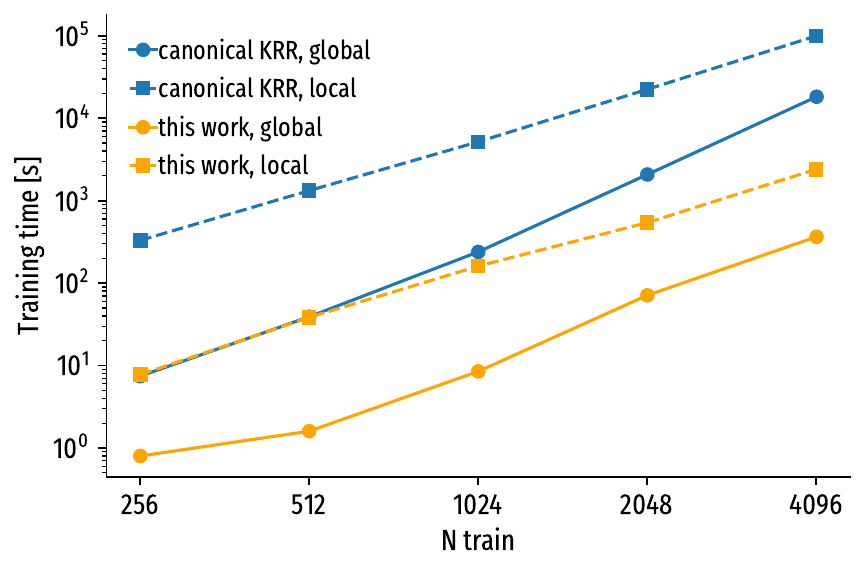}
    \caption{Training time as function of the training set size. In blue, models trained with a canonical KRR approach; in orange, models trained using the models mentioned above. The models where trained on cc-pV5Z data using the local and global versions of SLATM. All models where trained on a  AMD EPYC 7513 using a single core.}
    \label{fig:time_comparison}
\end{figure}

The models in Fig~\ref{fig:time_comparison} were trained with a grid of $21$ length scales $\sigma$ and $10$ regularization strengths $\lambda$  with $50$ shuffles. The total time $t$ needed to build such a model is:
\begin{equation}
    t= t_\text{D}+n_\sigma t_\text{K}+n_\text{s}n_\sigma n_\lambda t_\alpha
    \label{eqn:timing}
\end{equation}
Here $t_\text{D}$ and $t_\text{K}$ are the times to build the pairwise distance and the kernel matrix, while $t_\alpha$ is the time to solve \ref{eqn:krr}, $n_\sigma,\,\,n_\lambda$ and $n_\text{s}$ are the number of the corresponding hyperparameter and shuffles used in cross-validation. While this expression is in theory true for both global and local descriptors, there are some limiting factors. For local descriptors the size of the pairwise distance matrix and the kernel matrix are too large for most systems to keep them in memory at the same time, e.g. for 4096 training points both matrices would each require $53\,\text{GB}$ additionally the local descriptors would require around $17\,\text{GB}$. Therefore, it is necessary to recalculate both $t_\text{D}$ and $t_\text{k}$ for each $\sigma$. Using the approximation of the Chebyshev polynomials for local kernels we can save this costly step.

By applying all the techniques described in the method part, we can observe a performance increase of more than one order of magnitude in comparison to the canonical version of KRR. In the case of the global descriptor the time for 4096 data points is reduced to $17.5\%$ of the original cost, while the cost for the local version is reduced to $2.4\%$ of the canonical version for the same train size. However in this comparison we assumed that we can use the decompositions and the approximation for every pair of hyperparameter on this data set, this is not formally guaranteed for every data set and might lead to different performance curves depending on the data set and hyperparameter grid.
Both Eigen and Schur decomposition have a direct impact on $t_\alpha$ while the Chebyshev approximation only improves the time $t_\text{K}$ for local descriptors. In equation~\ref{eqn:timing} we can see that $t$ scales with the size of the hyperparameter grid and the number of shuffles, as a result our method becomes more valuable for larger hyperparameter scans.
\subsection{Learning the CBS Limit}
The most straightforward strategy is direct learning of the raw energy convergence towards the cc-pV5Z target. This means starting from pcseg-0  without any additional processing. For the GDB-BSIE dataset, a similar approach has been considered starting from STO-3G\cite{Holm2023}.

We observed the residuals to have a high variance, with a standard deviation of  1.1 Ha, indicating that the raw basis set incompleteness error (BSIE) is both large and irregular. This is in line with previous work\cite{Helgaker1997}, since S$\zeta$ basis set lies far from the continuous convergence asymptotic curve towards the CBS limit -- typically, even D$\zeta$ basis set results to not agree with the asymptotic scaling.  This baseline provides a useful reference point to understand the extent to which the model performance can be improved beyond this minimal direct-learning approach. Reducing the variance of this residual is key to improving overall learning efficiency, since a larger variance shifts the learning curve upwards.

\begin{figure}
    \centering
    \includegraphics[width=1\linewidth]{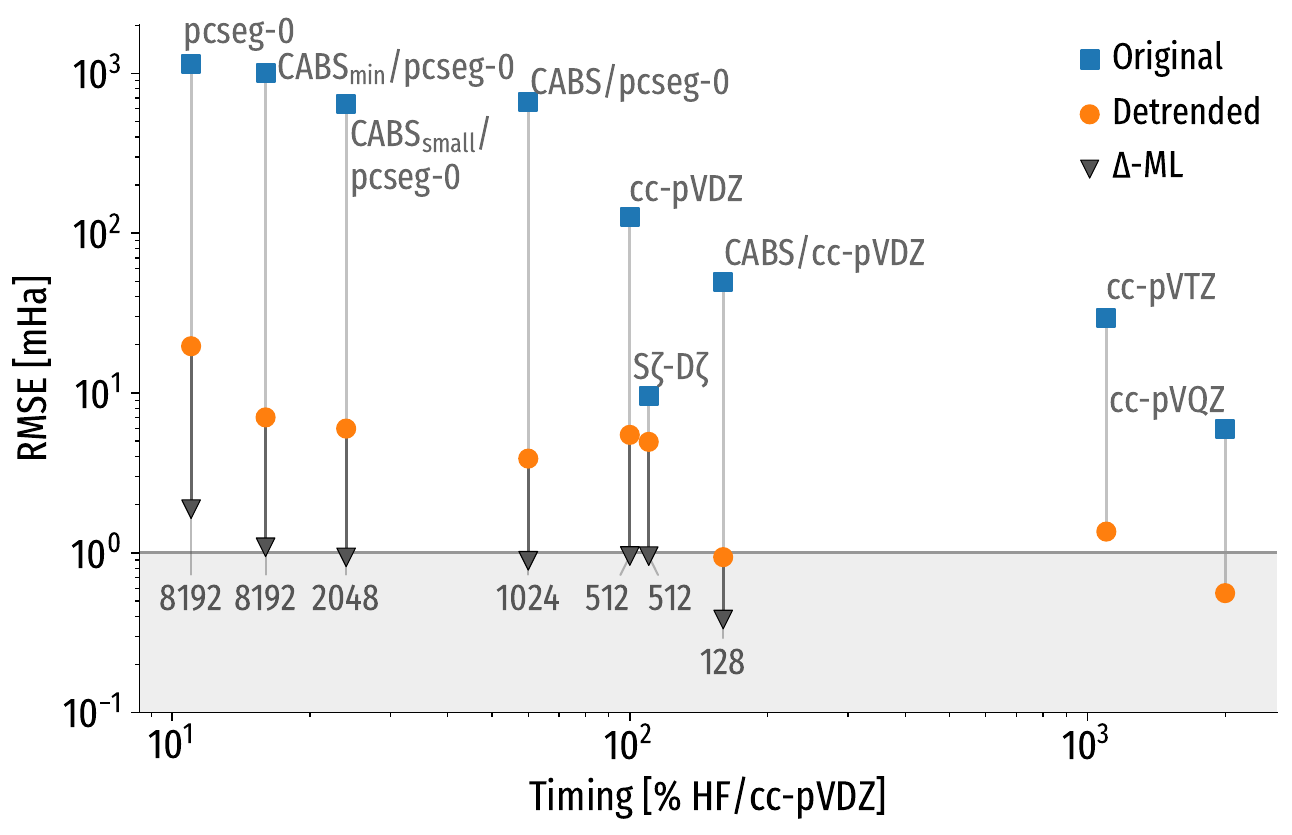}
    \caption{Residual errors as function of the computational cost of RHF/cc-pVDZ. In blue, the original values; in orange, the values after atomic detrending. The grey  points denote the results of a $\Delta$-ML approach with number of training points as annotations. See Table \ref{tab:rmse} for details. Target accuracy shaded.}
    \label{fig:RMSE}
\end{figure}

\begin{table}
    \centering
    \begin{tabular}{lcccc}\toprule
         Method&  Original&  Detrended&$\Delta$-ML (n)&Time\\\hline
         pcseg-0            &       1149.34&  19.62 &1.87&11\%\\
                     &       &  (98.3\%) &(8192) &\\
 CABS$_{min}$/pcseg-0&       1007.15&  7.03&1.08&16\%\\
                     &       &  (99.3\%) &(8192) &\\
 CABS$_{small}$/pcseg-0  &        647.04&  5.99 &0.94 &24\%\\
                     &       &  (99.3\%) & (2048)&\\
         CABS/pcseg-0        &         660.21&    3.89& 0.89&60\%\\
                     &       &  (99.4\%) &(1024) &\\
         cc-pVDZ&          126.29&    5.47 &0.95&100\%\\
                     &       &  (95.7\%) &(512) &\\
 Extrapolation/S$\zeta$-D$\zeta$&          9.59&  4.95 &0.95&110\%\\
                     &       &  (48.4\%)  &(512) &\\
         CABS/cc-pVDZ&            49.47&    0.94& &160\%\\
                     &       &  (98.1\%)  & &\\
         cc-pVTZ&            29.60&    1.36&  &1100\%\\
                     &       &  (95.4\%)  & &\\
         cc-pVQZ&           5.94&   0.56&  &2000\%
         \\
 & & (90.6\%)& &\\\end{tabular}
\caption{RMSE against the target energy (RHF/cc-pV5Z) in mHa, shown for raw data, atomic detrending (in parentheses: relative error reduction), $\Delta$ Machine Learning (in parentheses: number of training points required to reach 1\,mHa RMSE) together with the \textit{inference}/query time relative to a HF/cc-pVDZ single point.}
\label{tab:rmse}
\end{table}
With the idea of pushing further down the reduction the value of the residuals, we consider another post-processing approach: an atomic decomposition scheme. This approach partitions molecular properties into per-atom contributions based on physical considerations or data-driven criteria. Atomic detrending identifies the baseline energy contribution of individual atoms by representing a model based on chemical composition. By substracting this "trend" from the raw data, it is possible to isolate the residual energy $\epsilon$ that cannot be captured by the purely additive atomic model 
\begin{equation}
    E \approx \sum_{Z} n_{Z} c_{Z} + \varepsilon
    \label{eq:placeholder_label}
\end{equation}
where $n_Z$ is the number of atoms of element $Z$ in the molecule, $c_Z$ are fitted atomic coefficients obtained via linear least squares, and $\varepsilon$ is the residual that cannot be captured by a purely additive atomic model. This simple decomposition does not require any additional computational cost, regardless the computational method.

To calculate the CBS limit energy, the extrapolations based on two or three (consecutive) cardinal points are probably the most established approaches. Although the goal of this work is to be able to achieve CBS accuracy set at cc-pV5Z with the smallest possible basis set and computational cost, we build an extrapolation using the S$\zeta$ and D$\zeta$ basis sets for comparison despite D$\zeta$ being about nine times more expensive as pcseg-0. Since the parameters for extrapolation are validated for higher cardinal numbers (D$\zeta$/T$\zeta$ and T$\zeta$/Q$\zeta$) only, we first needed to obtain the corresponding values for this pair of smaller basis sets. 

\begin{figure}
    \centering
    \includegraphics[width=1\linewidth]{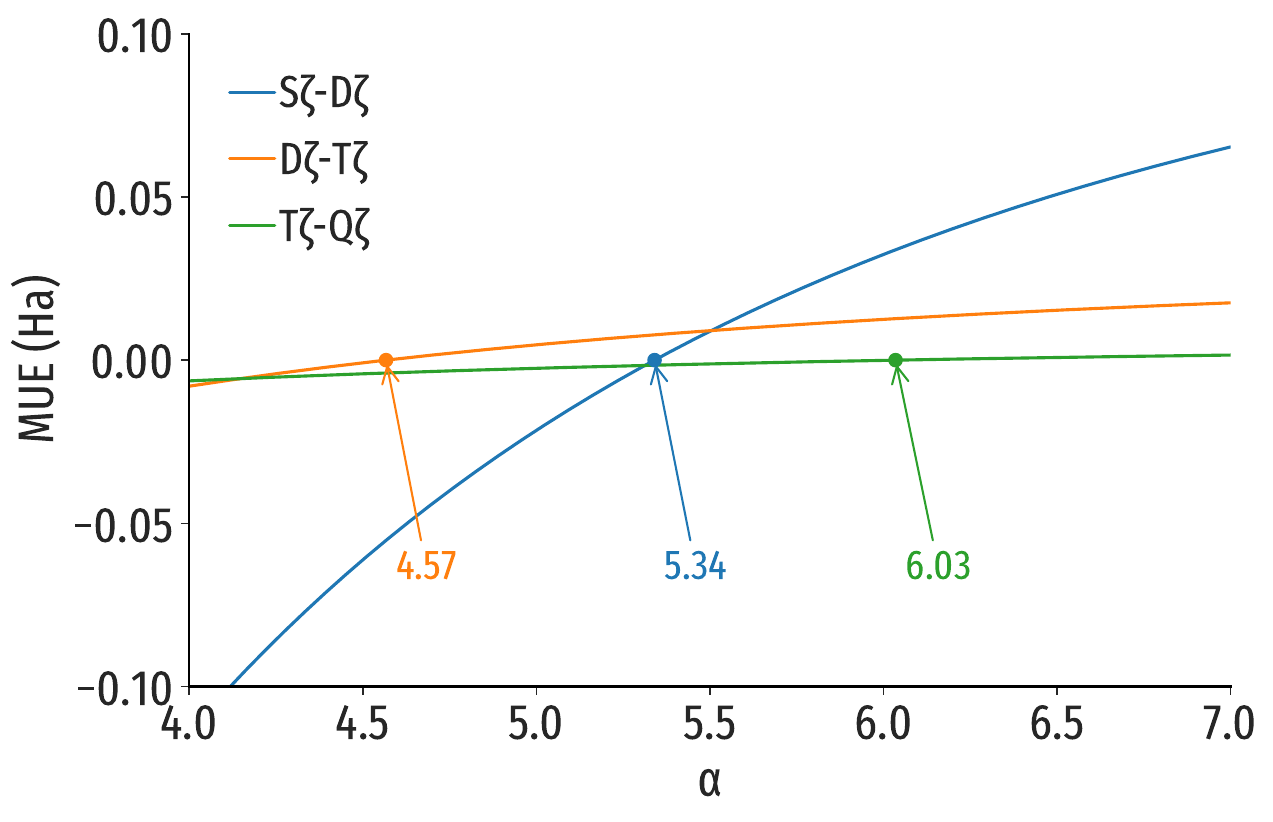}
\caption{Screening of $\alpha$ values for the extrapolations of S$\zeta$/D$\zeta$ (blue), D$\zeta$/T$\zeta$ (orange), and T$\zeta$/Q$\zeta$ (green) vs MUE error for 131\,554 points}
    \label{fig:alpha_curve}
\end{figure}

The dataset considered in this work, does not permit calculation of both decay coefficients $\alpha$ (for mean field) and $\beta$ (for correlation), since the correlation energies are not available. For this reason, only the $\alpha$ value is calculated based on eqn. \ref{eq:neese}. We estimated the $\alpha$ value that allows us to calculate the S$\zeta$/D$\zeta$ extrapolation employing the calculated RHF/pcseg-0 energies together with the RHF/cc-pVDZ retrieved from the dataset. Using $131\,554$ data points available in the GDB-BSIE dataset, we obtained an optimal $\alpha$ value of 5.34 (Figure~\ref{fig:alpha_curve}). The calculation of the $\alpha$ value for the remaining cc-pVXZ family pairs (D$\zeta$/T$\zeta$ and T$\zeta$/Q$\zeta$) yields values of 4.57 and 6.03, respectively. These values slightly differ from those found by Neese and Valeev for the same pairs (4.42 and 5.46), which can be partially attributed to a larger dataset with bigger molecules and a lower target energy (5Z vs 6Z). 

For the S$\zeta$/D$\zeta$ extrapolation towards the cc-pV5Z reference, the detrended error of the residuals is highly reduced to a single digit, 4.95 mHa. This value is close to the one that can be obtained using only cc-pVDZ with atomic detrending applied (5.47 mHa). This is a reflection of the low effect of the atomic detrending for this method in comparison with all other methods studied (less than 50\% against over 90\%). This means that calculating the two point extrapolation, which requires 10 times that of the S$\zeta$, provides negligible improvement to simple D$\zeta$, which already requires about nine times the computational cost of S$\zeta$. 
For this scenario, the atomic contribution is already accounting for part of extrapolation improvement, which suggests that this two-point scheme mainly corrects the contribution of individual atoms. Instead, we can investigate the CABS correction (see above). In Psi4, this is implemented\cite{Mitchell2024} as part of MP2-F12.

At the heart of the CABS singles correction is a basis set which is not used in the Hartree-Fock calculation at all, but comes in in a post-processing step where it is supposed to cover the remaining degrees of freedom that are not described by the original basis set (OBS). Besides manual optimisation\cite{Yousaf2008,Hill2014}, automated methods have been developed. We use AutoCABS\cite{Semidalas2022}, which naturally has been tested on D$\zeta$ and larger so far. Using PSI4 1.10 we retrieve the approximate timing required for this CABS, and observe that this additional step (dominated by the Fock matrix build) adds up to a total of approximately 60\% to the calculation of cc-pVDZ or renders pcseg-0 roughly as expensive as the DZ calculation on its own, and as such adds no direct computational benefit over DZ out-of-the box.

\begin{figure*}[t]
    \centering
    \includegraphics[width=1\linewidth]{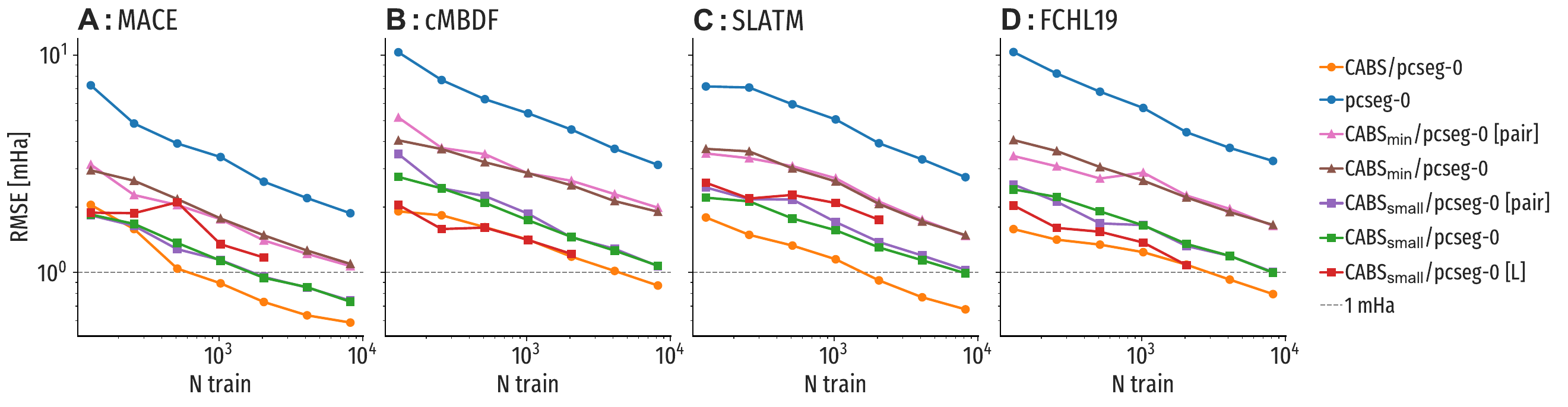}
    \caption{Comparative $\Delta$-ML learning curves of all pcseg-0 approaches with the 4 selected descriptors to study in this work. Results are shown for (A) MACE, (B) cMBDF, (C) SLATM, and (D) FCHL19. Notation: [L] = local descriptor with elemental masking; [pair] = gCP detrending applied.}
    \label{fig:learning_curves}
\end{figure*}

However, incorporating the full RHF-CABS correction with the CABS obtained from AutoCABS package into the pcseg-0 energy and a follow atomic detrending, provides a RMSE error which is even lower than the corresponding to cc-pVDZ with atomic detrending (3.89 vs 5.47\,mHa). This means that, with six times the cost of a pcseg-0 calculation (instead of nine times), we achieve an error lower than 4 mHa (which is lower than that obtained for RHF/cc-pVQZ without post-processing).

Based on the CABS optimization study described above, we set two additional approaches to include the complementary basis sets to improve simple pcseg-0 while pushing the time as minimum as possible; we defined the minimal approach (RHF-CABS$_\textrm{min}$/pcseg-0), using only D functions, since their introduction to the CABS reduced the RMSE by half, and a slightly larger one, called small (RHF-CABS$_\textrm{small}$/pcseg-0), which is made from PPDD functions. This version with additional functions presented the lowest RMSE of the optimized CABS.

The first one, which requires only 5\% extra computational cost over pcseg-0, relative to D$\zeta$, provides barely a reduction of the error by a 10\%. The second approach (RHF-CABS$_\textrm{small}$) is more efficient in the non-detrended case: It requires only twice the time of a basic RHF/pcseg-0, achieving the same error as the complete RHF-CABS/pcseg-0, which requires six times the compute time of the S$\zeta$. This means that the functions of the CABS can be optimized per basis-set family. In this case, they can simply be described with the PPDD functions, as defined in the CABS of RHF-CABS$_\textrm{small}$. 

With atomic detrending, the picture changes substantially: for these three CABS variants, the RMSE reduces to a single digit: between 7.03 mHa for the RHF-CABS$_\textrm{min}$ to 3.89 mHa for the full RHF-CABS. This suggests that despite the size and amount of functions used to describe the RHF-CABS methods, this technique reduces the error to less than 1\%. From this analysis, it can be assumed that any RHF-CABS model captures big part of the energy that cannot be described by the atomic contribution. 

The analysis of  these results, and those of the subsequent cardinals T$\zeta$ and Q$\zeta$ levels, showcase that using only the post-processing technique of atomic detrending to any cardinal number, a similar RMSE to the non-detrended one after next cardinal number is achieved, adding little extra computational cost. Removing the atomic contribution for these high cardinal numbers, provides nearly chemical accuracy for cc-pVTZ (1.36\,mHa) while successfully reaches chemical accuracy for cc-pVQZ.
The use of both of this post-processing techniques, yields smaller and more homogeneous  residuals, which in turn provide an improved target for ML models.

Using our AutoKRR package, we test four different descriptors: SLATM, cMBDF, FCHL19 and MACE for all the methods that would match or decrease state of the art speed (D$\zeta$), while increasing the accuracy fig (\ref{fig:learning_curves}). The target accuracy was set at a basis set incompleteness RMSE error of 1\,mHa.

We starting by evaluating the global versions of the four descriptors. The first observation that caught our attention was that MACE descriptors are the most data-efficient across all groups, consistently requiring fewer training points than the remaining descriptors. While, on the other hand, cMBDF showed to be the descriptor that require the most amount of training points to reach the target accuracy throughout all different models. It is the only one that fails to reach target accuracy for $\Delta$ML of RHF-CABS$_\textrm{small}$ (within the limit of 8192 evaluated points).
SLATM and FCHL19 behave similarly and often interchangeable. Of all of these methods, it is interesting to highlight that using only 2048 points an accuracy below 1 mHa (0.94 mHa) is achieved with RHF-CABS$_{small}$/pcseg-0 and MACE descriptor. This method requires only 24\% of RHF/cc-pVDZ time, which, on the other hand, require 512 or less points. 

Based on the results of the optimization of the CABS (Figure~\ref{fig:cabsopt}), we were interested in studying the pair detrending, since they present lower RMSE. This result was observed for both, the minimal and the small RHF-CABS approaches, so we studied them employing gCP. In this case we included the minimal since it was defined based on pair detrending (Figure~\ref{fig:cabsopt}) and we wanted to check if it had any additional effect. In any of both case of RHF-CABS$_\textrm{min}$/pcseg-0, there is no difference in the learning curves regardless the detrending approach used for the model training, beyond the null model. 
As well as the study of global descriptors, we study local descriptors including elemental masking. It is worth noting that since local descriptors requires higher computational resources and practicality, this study was run exclusively for RHF-CABS$_\textrm{small}$/pcseg-0, since it is the smallest CABS model that achieves the target accuracy. The study was made for all four descriptors and up to 2048 points, since that was the amount of points in which the threshold was reached with the global descriptor. Interestingly, for both cMBDF and FCHL19 descriptors, the local approach provides better results than the global one, behaving similarly to the next RHF-CABS system (RHF-CABS/pcseg-0) model. On the other hand, for MACE and SLATM the opposite occurs. The learning curve for the this local descriptors lays between the RHF-CABS$_\textrm{min}$/pcseg-0 and RHF-CABS$_\textrm{small}$/pcseg-0, performing worse than the global approach.

\section{Conclusions}
In this work, we showed that the complete-basis-set (CBS) limit Hartree–Fock energies can be calculated fast and efficiently by combining quantum-chemistry motivated corrections with data-driven learning, rather than classic extrapolation approaches or expensive basis sets. We showed that the BSIE variance can be heavily reduced by simple post-processing strategies, such as CABS and, especially, atomic detrending, making the residuals smaller and easier to learn. This indicates that the BSIE has a systematic structure that can be captured before applying a machine learning approach.
We showed that, using inexpensive HF/pcseg-0 calculations augmented with optimised physical corrections and $\Delta$-learning with only 2048 training points, we can recover near-CBS accuracy at a small fraction of the usual computational cost (24\% of a D$\zeta$ calculation). In particular, we found that different F12/CABS variants, such as RHF-CABS$_{small}$/pcseg-0, provide a highly favourable cost-accuracy trade-off compared to established approaches. 
Using show that standard Kernel-Ridge-Regression approaches can be accelerated by almost two orders of magnitude using Chebyshev polynomials, the Eigen decomposition and the Schur complement. Our implementation, AutoKRR, openly implements these methods and performs full cross-validation and hyperparameter optimisation of a model with 1000 training points in a few seconds (global kernel) or two minutes (local kernel) on a single core with modest memory requirements, which makes the methods substantially easier to apply.
When we learned these significantly reduced residuals with kernel ridge regression, especially using MACE descriptors, we were able to achieve a BSIE below 1 mHa, using a few thousand points at less than 25\% of the cost of a cc-pVDZ calculation. 
The results are relevant for those quantum chemistry calculations where the CBS limit of Hartree-Fock needs to be estimated, e.g. for quantum alchemy. Overall, we demonstrate a practical, inexpensive and scalable CBS estimation approach.

\section{Data Availability}
Our custom implementation of CABS is available in the Python package \texttt{nablachem.cabs}. All code regarding the Kernel Ridge Regression is available as the Python package \texttt{nablachem.krr}.

\section{acknowledgments}
The authors thank Konstantin Karandashev and Werner Seiler for the discussions.

\bibliography{nablachem}
\end{document}